# Bias Smells in AI Software Development: Recognizing Potential Sources of Fairness Debt

RONNIE DE SOUZA SANTOS, University of Calgary, Canada
CLEYTON MAGALHÃES, Universidade Federal Rural de Pernambuco (UFRPE), Brazil
RODRIGO SPINOLA, Virginia Commonwealth University, USA

**Context:** Fairness debt arises from AI development shortcomings that may lead to societal harms. While technical and social debt concern software design decisions and team dynamics, respectively, fairness debt captures the long-term consequences of development decisions that may reinforce bias and inequities in AI-based software systems. Despite growing attention to AI fairness, limited evidence exists on how practitioners recognize potential sources of fairness debt during development. **Aim:** This study investigates the indicators practitioners recognize as signaling potential sources of fairness debt in AI-based software projects. **Method:** We conducted an exploratory case study of four AI projects within one organization. Data were collected from 25 professionals through semi-structured interviews and open-ended questionnaires, complemented by observation of internal communication channels and project documentation, and analyzed using iterative qualitative coding, memoing, and constant comparison. **Results:** We identified six recurring indicators, termed bias smells: Context Oversimplification, Dataset Imbalance, Metrics Inadequacies, Ad hoc Testing, Individual Diversity Unawareness, and Homogeneous Team Composition. These smells span technical and human aspects of software development and signal conditions that may introduce or reinforce bias and contribute to fairness debt. **Conclusion:** Bias smells extend the software smell paradigm to fairness and provide a foundation for incorporating fairness into software quality assurance through observable indicators.

## 1 Introduction

The quality of software relies on a well-structured development process that intertwines technical factors with decision making [15]. In this context, technical debt arises when developers take shortcuts to accelerate software delivery, often at the cost of long-term maintainability [8, 30, 51]. These shortcuts, while expedient, accumulate as hidden costs, increasing the complexity and effort required for future modifications. Code smells serve as indicators of technical debt, highlighting problematic areas in the codebase that are likely to degrade maintainability over time [14, 22, 44]. They provide observable indicators that can support software quality assurance by drawing attention to potential maintainability problems before they become more costly to address.

Authors' Contact Information: Ronnie de Souza Santos, ronnie.desouzasantos@ucalgary.ca, University of Calgary, Calgary, Alberta, Canada; Cleyton Magalhães, cleyton.vanut@ufrpe.br, Universidade Federal Rural de Pernambuco (UFRPE), Recife, Brazil; Rodrigo Spinola, spinolaro@vcu.edu, Virginia Commonwealth University, VCU College of Engineering, Richmond, VA, USA.

Social factors also play an essential role in software development. In this context, social debt emerges when poor team dynamics, inadequate communication, and unresolved conflicts erode collaboration and morale [10, 14, 27, 28]. This debt accumulates over time, leading to decreased productivity, team dissatisfaction, and challenges in knowledge sharing. A key manifestation of social debt is community smells, which are patterns of dysfunction in software teams, such as conflicts, siloed work structures, or lack of engagement, that disrupt team cohesion and efficiency [11, 20]. Community smells thus provide observable socio-technical indicators that can help organizations recognize collaboration problems and extend quality-oriented monitoring beyond the source code itself.

More recently, with the rise of AI-enabled systems and their significant societal impact, the concept of fairness debt has emerged [37–39]. Unlike technical debt, which primarily concerns software maintenance and evolution, and social debt, which concerns problematic socio-technical conditions within development communities, fairness debt captures the long-term consequences of development decisions and practices that may contribute to discriminatory outcomes or inequitable treatment of individuals or groups [18, 37, 38]. Fairness debt may arise from design, implementation, managerial, or data-related decisions and practices that leave fairness concerns insufficiently addressed, allowing their consequences to persist and making subsequent mitigation increasingly difficult or costly. Such consequences may extend beyond the software development environment, contributing to social inequalities, legal challenges, loss of trust, and other societal harms [37–39].

While technical debt and social debt are well-established concepts characterized by observable indicators, such as code smells [8, 14, 22, 30, 44, 51] and community smells [11, 20], fairness debt remains a relatively recent concept whose indicators have received limited empirical attention [37, 38]. Existing AI fairness research has largely focused on detecting, measuring, and mitigating bias in datasets, models, and system outcomes. Comparatively less attention has been given to development decisions, practices, and conditions that may introduce or reinforce fairness-related risks, and to the indicators practitioners use to recognize such conditions during software development. Understanding these indicators is important for AI software quality assurance because they may enable development teams to recognize potential sources of fairness debt before their consequences become more difficult or costly to address.

Inspired by the use of code smells and community smells as indicators of technical and social debt, this study examines the recurring signs that practitioners associate with potential sources of fairness debt in AI-enabled software projects. Drawing on practitioners' experiences in real-world development settings, we characterize socio-technical conditions that may introduce or reinforce bias and thereby contribute to fairness debt. We refer to these recurring indicators as *Bias Smells*, extending the software smell paradigm to fairness-oriented software quality assurance in AI-enabled systems. Accordingly, we investigate the following research question:

> **RQ: What indicators do practitioners recognize as signaling potential sources of fairness debt in AI-enabled software projects?**

By addressing this research question, this study makes the following contributions:

- Proposes the concept of *Bias Smells*, extending the software smell paradigm to characterize observable indicators of potential sources of fairness debt in AI enabled software systems.
- Identifies six recurring indicators that practitioners associate with potential sources of fairness debt during the development of AI enabled software systems.
- Organizes these indicators into a socio technical perspective comprising technical and human related indicators, providing a structured understanding of how fairness debt may emerge throughout software development.

- Discusses how these indicators are related to software quality assurance for AI enabled systems, supporting the early identification and mitigation of fairness related risks.

The remainder of this paper is organized as follows. Section 2 presents the background supporting this work. Section 3 describes the research method. Section 4 presents the findings. Section 5 discusses the results, and Section 6 presents their implications for research, practice, and society. Finally, Section 7 concludes the paper and summarizes the main contributions.

## 2 Background

In this section, we discuss the three main concepts addressed in this study: technical debt, social debt, and fairness debt.

### 2.1 Technical Debt and Code Smells

Technical debt refers to the accumulation of future costs resulting from expedient but suboptimal design choices or coding practices in software development [7, 33]. These choices often stem from the pressures of balancing competing constraints, such as limited budgets, strict release deadlines, and evolving software requirements. While these shortcuts may accelerate initial development, they introduce hidden deficiencies in the codebase that degrade software maintainability, flexibility, and extensibility over time [34]. The metaphor of "debt" is apt because, much like financial debt, technical debt incurs interest, measured by the extra effort, cost, and risk required for future modifications and improvements [7].

One of the primary indicators of technical debt is the presence of code smells, which are patterns in the codebase that suggest deeper design flaws or maintenance challenges [36]. These smells do not necessarily indicate immediate defects, but they signal areas where the code is likely to become harder to maintain, extend, or debug over time. Common examples of code smells include duplicated code, where identical or highly similar code appears in multiple locations, increasing the risk of inconsistencies when changes are made; long methods, which perform too many tasks and become challenging to understand and modify; and large classes, which accumulate excessive responsibilities, violating the Single Responsibility Principle and complicating refactoring efforts. Additionally, complex conditional logic, characterized by deeply nested or convoluted conditional statements, reduces code readability and increases the likelihood of errors, while god classes, i.e., classes that centralize excessive functionality, create tightly coupled dependencies that hinder modifications and system flexibility [22].

These smells often emerge due to rushed development, inadequate refactoring, or violations of fundamental software design principles [14, 22, 44]. While some level of technical debt is unavoidable in real world projects, teams that fail to address these issues systematically risk accruing unsustainable amounts of debt, which jeopardizes software quality and long term project success. Recognizing and mitigating technical debt through continuous refactoring, adherence to coding standards, and the use of automated tools for code analysis is essential for sustaining maintainable and scalable software systems [24, 40].

### 2.2 Social Debt and Community Smells

Social debt refers to the accumulation of negative consequences resulting from suboptimal socio technical decisions within work environments [11]. Unlike technical debt, which primarily affects software artifacts, social debt manifests in the interpersonal and organizational dynamics of development teams, leading to strained relationships, communication breakdowns, and reduced productivity. This form of debt arises when poor social decisions, such as ineffective communication practices, exclusionary team structures, or unresolved conflicts, persist over time and amplify their detrimental effects. Just as technical debt incurs interest in the form of higher maintenance costs, social debt grows when these social issues remain unaddressed, leading to reduced morale, increased turnover, and inefficiencies in collaboration [12, 17, 46].

A key manifestation of social debt is the presence of community smells, which are patterns of dysfunction that emerge in software teams due to inadequate socio-technical interactions [9, 11, 12, 17, 46]. Prior work by Tamburri et al. [11], Caballero et al. [15], and others identifies 30 recurring community smells, categorized by causes and effects. Causes include rigid hierarchies, siloed communication, or power imbalances, while effects include strained interactions, exclusionary behaviors, and inefficient knowledge sharing, which weakens team cohesion and the overall development process.

Common examples of community smells illustrate how social debt accumulates in software teams. The Organizational Silo Effect arises when subgroups operate in isolation, which limits collaboration and communication. The Black Cloud Effect occurs when developers lack access to essential project knowledge, leading to uncertainty and inefficiencies. The Prima Donnas Effect surfaces when dominant individuals reject contributions from others, which creates a hierarchical and unwelcoming environment. Sharing Villainy reflects poor information exchange, which reduces transparency and complicates decision making. The Organizational Skirmish smell results from mismatched expertise levels and communication channels, which causes friction in workflows. The Lone Wolf effect occurs when isolated individuals hold critical knowledge, which creates bottlenecks and makes the team overly dependent on them. Another example is Architecture by Osmosis, where architectural decisions are made informally without proper documentation or knowledge management, leaving key information restricted to a group of individuals, while others lack access to important design choices, which leads to inefficiencies, repeated discussions, and unstable system configurations [15, 46].

Just as refactoring and automated code analysis help manage technical debt, social interventions, such as regular feedback sessions, mentorship programs, and conflict resolution strategies, can help maintain a healthy, collaborative development environment. By recognizing and addressing community smells early, software teams can prevent long term negative consequences, improve work culture, and enhance overall software quality [15, 23, 48].

### 2.3 Software Fairness and Fairness Debt

Software fairness ensures that software systems, algorithms, and their outcomes remain ethical, equitable, and unbiased across different groups of people [4, 6, 42, 52]. This is particularly important in AI-driven systems, where biases embedded in data and decision-making processes can create widespread societal harm. Unlike technical and social debt, which manifest within the software development environment and affect code, design, and team dynamics, fairness debt extends beyond software projects and influences broader societal structures [37, 39]. Fairness debt accumulates when biased design, implementation, or managerial decisions create long-term challenges in achieving justice and equity in algorithmic outcomes. This form of debt can disproportionately affect marginalized communities, reinforcing discrimination and restricting access to essential services and opportunities [38, 39].

Fairness debt can stem from multiple biases, including cognitive, historical, model, requirement, societal, and training biases, all of which shape how software systems process and interpret information [18, 39]. Unlike technical debt, which leads to inefficiencies in code, or social debt, which disrupts teamwork and collaboration, fairness debt can affect individuals outside the development environment. Additionally, fairness debt may be more difficult to address once systems are deployed, since its effects can become embedded in broader institutional and societal contexts, which may require interventions beyond software refactoring or internal process improvements [37, 39].

For example, biased credit scoring models can disadvantage lower-income applicants, making it harder for them to access financial services. In healthcare, AI-driven diagnostic tools trained on nondiverse datasets may lead to misdiagnoses for underrepresented populations, affecting the quality of care they receive. Similarly, social media algorithms designed to maximize engagement may amplify harmful stereotypes or suppress certain viewpoints, shaping public discourse in ways that reinforce discrimination. In such cases, fairness-related concerns may not have been sufficiently identified and addressed during development, allowing inequities to persist in deployed systems [18, 39].

Reducing fairness debt requires active steps, such as fairness-aware software development, the use of diverse and representative datasets, and ongoing audits of algorithmic decisions to identify and mitigate biases [37, 39]. While technical and social debt can often be managed within software teams, fairness debt may require engagement with policymakers, domain experts, and affected communities to support more accountable AI systems. Incorporating fairness considerations throughout the development lifecycle may help limit the accumulation of such debt and its broader impacts.

### 2.4 Comparing Technical, Social, and Fairness Debt

Technical, social, and fairness debt represent distinct forms of long term liabilities that emerge from decisions made during software development. Although they differ in their causes and consequences, they share a common characteristic: they accumulate over time as a result of suboptimal decisions whose immediate benefits come at the expense of future costs. Their primary distinction lies in the entities they affect, the consequences they produce, and the observable indicators practitioners use to recognize them.

Technical debt concerns software artifacts. It results from design and implementation decisions that simplify short term development but increase future maintenance effort, architectural complexity, and defect proneness [33, 36, 47]. Practitioners recognize potential technical debt through observable indicators known as *code smells*, such as duplicated code, large classes, long methods, excessive coupling, or overly complex conditional logic. These indicators reveal underlying design deficiencies and motivate refactoring activities before software quality deteriorates further.

Social debt concerns the human and organizational aspects of software development. It emerges from socio technical decisions that negatively affect communication, collaboration, and knowledge sharing within development teams [9, 11, 12, 17, 46]. Similar to technical debt, practitioners recognize potential social debt through observable indicators, namely *community smells*. These recurring patterns of dysfunctional collaboration reveal socio-technical issues that may compromise team effectiveness and project sustainability.

Fairness debt concerns the societal consequences of software development decisions that may lead to discriminatory or inequitable outcomes once AI-enabled systems are deployed. It may originate from decisions involving datasets, requirements, model design, implementation, testing, or organizational practices. Although substantial research has investigated algorithmic bias, fairness metrics, dataset quality, and ethical principles, comparatively little attention has been given to how practitioners recognize early signs that fairness debt may be accumulating. Unlike technical debt and social debt, fairness debt currently lacks practitioner-oriented indicators analogous to code smells or community smells. To address this gap, we introduce the concept of *Bias Smells*.

## 3 Method

We conducted an exploratory case study [41] to investigate how software teams develop AI-based solutions and how indicators of biases emerge in practice. An exploratory case study is an empirical method that investigates a contemporary phenomenon in depth and within its real-world context, particularly when the boundaries between the phenomenon and the context are not clearly defined [41]. This design is appropriate for studying socio-technical phenomena in software engineering, where technical, organizational, and human factors are intertwined [35, 41].

Following the case study methodology [35], interviews served as the primary source of evidence, with additional sources such as observations and document analysis used to characterize how bias emerges and could accumulate in the software project, similar to what happens to code smells or community smells [15, 25]. Within this case study, we used elements of grounded theory [5] with a data analysis approach planned to systematically generate concepts grounded in empirical material. The grounded theory elements supported iterative cycles of data collection and analysis that involved inductive coding, memoing, and constant comparison [21]. This combination of case study design and inductive analysis enabled an in-depth investigation of the real-world setting while allowing concepts to emerge from practitioners' experiences.

### 3.1 Software Development Context

The empirical setting for this case study is a large software services company, headquartered in Brazil with additional operations in Portugal and the United States. The company delivers digital solutions to clients in finance, telecommunications, government, manufacturing, and services, serving markets across several continents. It employs roughly 1,200 people, over 70% of whom work directly in software development, distributed across dozens of teams that vary in technical composition, professional background, and preferred development approach. This site was chosen because it offered access to ongoing AI-based software projects and allowed direct observation of development practices in a real operational setting.

From this environment, four AI-based projects were selected for analysis, chosen on the basis of participant availability and organizational willingness to permit discussion of project work—that is, through convenience sampling rather than random or purposive selection. The four projects spanned distinct application domains, summarized as follows:

- **Project A** — A real-time Brazilian Sign Language translation system built on deep learning techniques [26], developed together with a Chinese multinational technology firm operating in more than 180 countries and maintaining a dedicated research and innovation hub. Work centered on constructing a video dataset of Brazilian Sign Language and building gesture-recognition modules suited to real-time, resource-constrained mobile deployment.
- **Project B** — A digital-twin and predictive-modeling initiative for a state-controlled oil and gas company with a strong footprint across Latin America, ranking among the largest firms of its kind worldwide. The effort targeted industrial process applications.
- **Project C** — An exploratory initiative applying large language models within a regional education setting, investigating their use in academic contexts.
- **Project D** — A computer-vision facial recognition system built in partnership with a major Brazilian cosmetics company that leads the Latin American market [1]. The resulting prototype detected lip contours and directed a robotic applicator to apply lipstick automatically, aiming to expand product accessibility for people with motor or visual impairments as part of a multi-year R&D collaboration.

Contractual constraints prevent us from disclosing further specifics about Projects B and C. Projects A and D, by contrast, have already been made public through separate reports [1, 26], and the descriptions offered here are calibrated to remain within these disclosure boundaries. Although the study is anchored in a single company, variation in application domain, client context, and AI technology across the four projects provided heterogeneous empirical contexts in which to examine how bias-related indicators emerge during AI development.

### 3.2 Participants

This case study draws on multiple participants involved in the selected projects, enabling the investigation of the phenomenon from different professional perspectives. In the selection process, we targeted designers, software engineers, data scientists, and testers engaged in the four projects described above. Participants were initially recruited using convenience sampling [3], based on availability and willingness to collaborate. The sample was then expanded through snowball sampling, where participants were asked to recommend colleagues who could provide additional perspectives [3]. This approach was particularly useful for reaching individuals from underrepresented groups who have specific views on biases and fairness and who might otherwise not have been included. To support this, participants were invited to share the study invitation within their networks without disclosing any personal information to the research team. In accordance with the ethics protocol, all participants contacted the researchers directly to express interest and opt in. As data collection progressed, we incorporated theoretical sampling [5] by inviting individuals with specific experience or demographic backgrounds not previously represented in the cohort. This allowed us to further develop emerging concepts and support a more comprehensive understanding of the phenomenon under investigation.

### 3.3 Data Collection

We collected data from one primary source (interviews) and two supporting sources (nonparticipant observation and document analysis). This multi source data collection strategy supports triangulation, allowing cross verification of findings from multiple perspectives and strengthening the study's credibility [35, 43]. The use of multiple data sources also supports the development of a chain of evidence, connecting observations to findings through consistent patterns across sources [43, 50]. In this study, the different sources provide complementary views of the same phenomenon: interviews capture participants' experiences and interpretations, observations provide insight into ongoing practices and interactions in context, and documents reflect formalized processes and artifacts. Therefore, interviews constituted the primary data, while observations and documents were used to contextualize and corroborate the findings. At the end, by systematically relating evidence from these sources during analysis, we ensure that findings are grounded in multiple forms of empirical material.

*3.3.1 Interviews.* We conducted semi structured interviews to capture participants' experiences with AI system development and bias. This approach balanced structure with flexibility, ensuring coverage of key topics while allowing participants to elaborate on their experiences. Before launching the study, we piloted the interview guide with two fairness researchers to test phrasing, clarity, and timing. Based on their feedback, we refined the script prior to the actual interviews. During the study, the guide was further refined iteratively by adding or rephrasing questions when new themes emerged [5]. Table 1 presents the interview guide.

We conducted interviews in three rounds to allow iterative refinement based on emerging insights and support participants context and experience. We began with seven participants per round and, by the third round, increased to eleven, at which point saturation was reached. The guide evolved over time: early rounds used predefined questions to explore sources of bias in AI development and challenges in addressing fairness; later rounds incorporated more open ended discussions on the indicators of lack of fairness (*smells*) so that practitioners could expand on prior themes, share examples, and provide contextual detail.

Between June 1 and July 5, 2024, we interviewed 25 professionals across roles that included software engineers, data scientists, designers, and testers. Interviews lasted from 17 to 35 minutes, generated around 6 hours of audio. To ensure flexibility, we accommodated participants' preferences for interview formats. While most interviews were conducted via video conferencing, three participants in round two were unable to participate in recorded sessions due to organizational restrictions. Instead, they completed a detailed open ended questionnaire, allowing them to share their perspectives in writing. Their responses were integrated into the analysis using the same coding approach.

Table 1. Interview Guide

| Interview Questions |
|---|
| **1. Can you describe your day-to-day work with the development of AI and machine learning systems?** |
| **Probe:** What are your main responsibilities in AI development? |
| **Probe:** What types of AI models or applications do you work on? |
| **Probe:** How does bias awareness factor into your work? |
| **Probe:** Have you encountered any fairness-related challenges in your projects? |
| **2. Algorithmic bias refers to distortions in machine learning outcomes that can lead to system errors, unfair decisions, or harmful consequences for specific groups—such as exclusion, discrimination, or reinforcement of societal inequalities. What strategies do you and your team use to identify algorithmic bias in your project?** |
| **Probe:** At what stage of development do you typically assess bias? |
| **Probe:** What indicators suggest that bias may be present? |
| **3. What indicators do you and your team observe that suggest a risk of bias before it becomes a problem?** |
| **Probe:** Are there warning signs during planning, requirements gathering, coding, testing, or data handling that suggest bias might emerge? |
| **Probe:** Are there specific decisions made early in development that later increase bias risk? |
| **4. What are suboptimal decisions (shortcuts or workarounds) made during AI development that help the system function in the short term but could contribute to bias or technical debt over time?** |
| **Probe:** What is a bad decision your team moved forward with that later ended up creating bias? |
| **Probe:** Have you encountered cases where technical debt or software smells contributed to bias in the system? |
| **5. I will now list some types of algorithmic bias, and I'd like you to describe whether you have observed them in your project and whether you or your team have made decisions that might contribute to them.** |
| - Cognitive Bias |
| - Design Bias |
| - Historical Bias |
| - Model Bias |
| - Requirements Bias |
| - Social Bias |
| - Testing Bias |
| - Training Bias |
| **6. Once you identify bias in your project, what steps do you and your team take to mitigate it?** |
| **Probe:** What processes or tools do you use to correct bias in models, data, or decision-making? |
| **Probe:** What challenges prevent your team from fully mitigating bias? |

*3.3.2 Nonparticipant Observation.* We conducted nonparticipant observation of online discussions within the organization to capture how practitioners discuss bias related concerns in their daily work. Within the company, two Slack groups were observed: Group 1 included 77 professionals discussing machine learning topics and Group 2 included 30 professionals discussing data science topics, with nine individuals participating in both groups. One researcher monitored discussions twice a week and produced field notes that recorded the content and tone of discussions, including contextual cues such as emojis and images when relevant. Observation was conducted during three periods: from June 1 to July 5, 2024, in parallel with the interviews; from March 15 to May 15, 2025, while the emerging findings were being refined; and again during February 2026 to support the consolidation of the analysis. These observations provided contextual data that complemented the interview findings and supported triangulation.

*3.3.3 Documents.* Document analysis served as another supporting source to validate findings through triangulation [29]. We explored testing documents related to Project B, including project descriptions, validation criteria, test cases, and internal team documents. These materials were examined throughout the study, from the beginning of data collection until early 2026, providing insight into formal development practices and allowing comparison with participants' accounts as the analysis evolved. Due to confidentiality, detailed replication packages cannot be released, but the materials were systematically analyzed as part of the study.

## 3.4 Data Analysis

The coding process followed three recommended steps in qualitative research [5, 43]: line-by-line coding, focused coding, and theoretical coding, moving from raw fragments of quotations extracted from interviews to conceptual categories and finally to higher-level concepts related to indicators of lack of fairness in AI development.

We started by transcribing the interviews and applied line-by-line open coding to the first round of data, identifying emerging concepts and their properties [5]. As new concepts emerged, we created memos to document early insights and track key themes. In the second round, we refined these codes through constant comparison, ensuring that each new data point was systematically compared against previously coded data. As interviews progressed, we integrated insights from non-participant observation, applying the same coding approach to online discussions. These observations provided real-world examples of how AI practitioners discuss and address bias-related issues in their daily work. We compared emerging themes across interviews and observations to identify consistencies and divergences.

To further enhance triangulation, we incorporated insights from document analysis, particularly testing reports and validation documentation from Project B. We analyzed these documents using focused coding to establish connections between bias-related indicators reported by participants and their documented practices. This triangulation between interviews, observations, and documents supported a clear chain of evidence linking data sources to findings.

By the final round of data analysis, recurring patterns related to indicators of lack of fairness had been identified, with findings reaching saturation [2]. At this stage, we conducted theoretical coding, refining relationships between categories to establish a cohesive narrative based on the experiences described by practitioners. Figure 1 illustrates the progression from raw data (quotations) through each coding phase.

## 3.5 Auditing

The first author led the initial coding process. The second author contributed to refining insights and validating coded data. The third author supported conceptual consistency. Given the diversity of contexts across the four projects and the interpretive nature of the analysis, we adopted a consensus based manual auditing process rather than agreement measures such as Cohen's kappa or Krippendorff's alpha. This approach is consistent with qualitative case study and other qualitative guidelines [35, 41], which emphasize discussion, reflexivity, and consensus over statistical agreement. All disagreements were resolved through consensus meetings, supporting rigor, transparency, and analytical credibility.

## 3.6 Researcher Positionality

The research team brought complementary perspectives to this study. The first author has extensive experience conducting qualitative research on fairness and human aspects of software engineering, particularly in teamwork and collaboration. The second author combines research on software development with practical industry experience in software engineering. The third author has extensive expertise in technical debt research. Furthermore, the first and third authors previously introduced and conceptualized the notion of *fairness debt*, which informed their familiarity with the phenomenon under investigation. We acknowledge that these prior experiences may have influenced data interpretation and the identification of emerging concepts. To mitigate this potential bias, coding decisions were documented through memos, interpretations were regularly discussed among the research team, and emerging findings were challenged during consensus meetings. This reflexive process is consistent with current recommendations for qualitative software engineering research and aims to improve the transparency and credibility of the analysis [13]

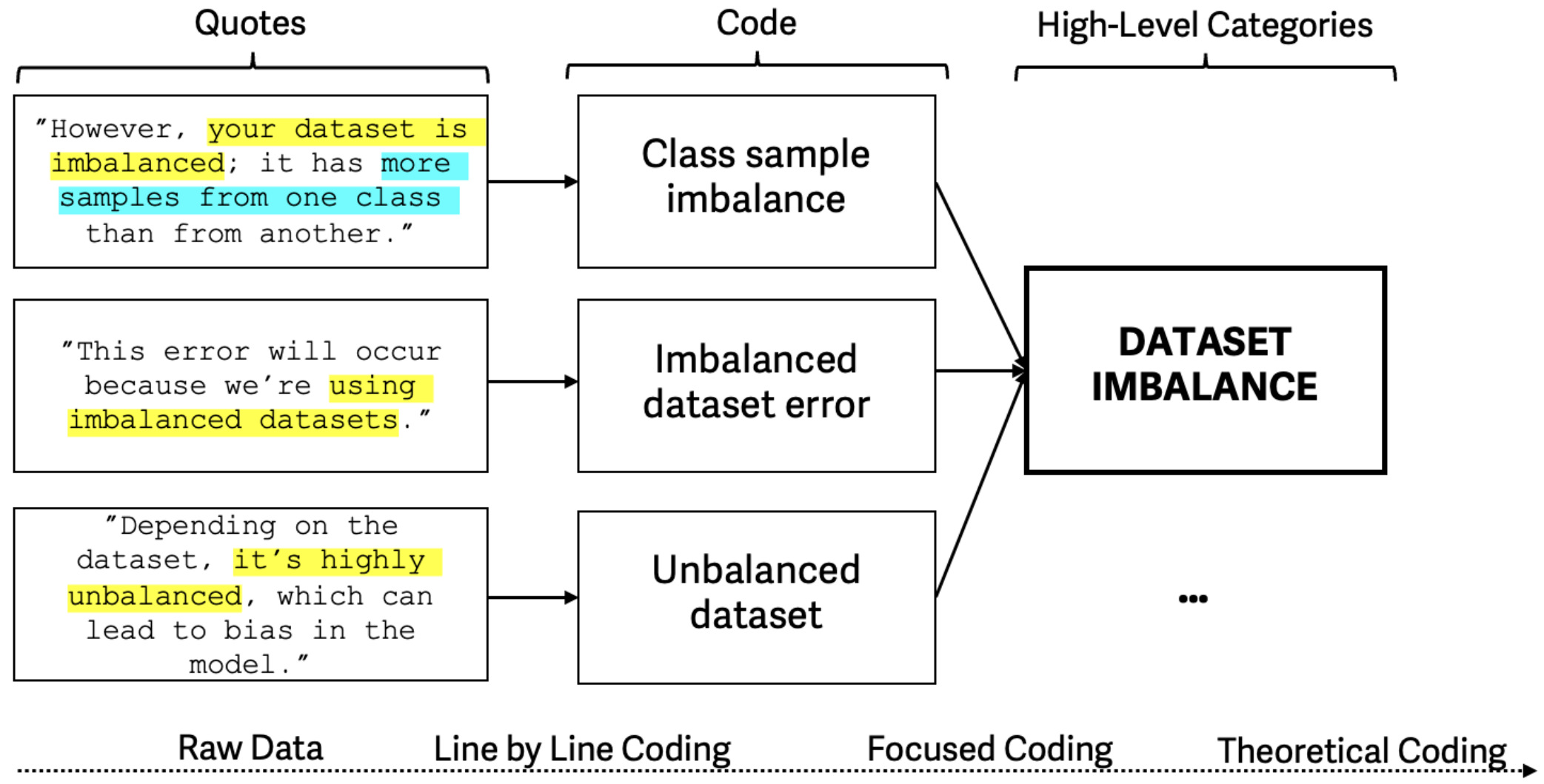

Fig. 1. Data Analysis Process

### 3.7 Ethical Considerations

We followed the ethics guidelines of the first author's university. The company permitted access to online discussions and documents. Participants were informed about the study, assured of confidentiality, and allowed to withdraw at any time. Due to contractual restrictions, the complete interview transcripts cannot be provided. However, sufficient methodological detail and representative quotations are reported to support analytical transparency and transferability.

### 3.8 Threats to Validity

As with any qualitative case study, our findings should be interpreted considering the characteristics of the research design. Rather than seeking statistical generalization, exploratory case studies aim to provide an in depth understanding of a phenomenon within its real world context and support theoretical rather than statistical generalization [35, 41, 50]. We therefore discuss the quality of this study in terms of credibility, reflexivity, and transferability. To enhance credibility, we established a clear chain of evidence linking observations to findings through systematic coding, memoing, constant comparison, and iterative refinement of emerging concepts. Interviews served as the primary source of evidence, while nonparticipant observations and document analysis were used to corroborate and contextualize the findings. This triangulation across data sources strengthened the consistency of the identified concepts and reduced reliance on a single perspective. Analytical credibility was further supported through consensus based auditing among the research team, where coding decisions and interpretations were discussed until agreement was reached.

Transferability was supported by providing a detailed description of the research context, the participating organization, the four investigated projects, participant roles, and the data collection and analysis procedures. Although the study is based on a single organization, the projects span different application domains, AI technologies, and client contexts, providing a richer basis for theoretical transfer to similar AI development settings. Nevertheless, the identified bias smells should not be interpreted as an exhaustive taxonomy nor assumed to be universally applicable. Researcher reflexivity was also considered throughout the study. Because members of the research team had prior experience investigating fairness debt and software engineering, coding decisions were documented through memos, interpretations were regularly challenged during consensus meetings, and multiple data sources were compared throughout the analysis. These practices helped reduce the influence of prior assumptions while maintaining transparency in the interpretation process.

Another potential threat concerns the time elapsed between the interviews and the completion of the analysis. The primary interviews were conducted between June and July 2024 and generated a substantial volume of qualitative material, requiring an extended period of iterative coding, memoing, and constant comparison. To reduce the risk that changes in organizational practices or AI technologies could affect the interpretation of the findings, we continued collecting supporting evidence through nonparticipant observations and document analysis during subsequent phases of the research, including additional observation periods in 2025 and 2026. These complementary sources were not used to modify participants' original accounts, but rather to contextualize, corroborate, and challenge the emerging interpretations, strengthening the credibility of the final findings. Finally, this study represents an initial empirical investigation of sources of bias in AI software development. Although the findings are grounded in evidence collected from four real AI projects, additional studies involving different organizations, domains, development processes, and AI technologies are needed to validate, refine, and expand our findings and to investigate their applicability across a broader range of software engineering contexts.

## 4 Results

In the section, we describe the categories that emerged from our analysis, which revealed the various types of indicators of bias (*smells*) observed by professional software developers when working on AI systems. We begin by presenting general information about the group of participants of the study, specifically those who took part in the interviews (Table 2). Following this, each indicator identified in our study is defined and described in detail, highlighting its characteristics and impact on AI development. We further illustrate each indicator using direct quotations from participants, showing how these elements manifest in real-world scenarios. Interpretive commentary is minimized in this section and developed separately in the discussion.

### 4.1 Demographics

As described in Section 3.2, we selected software professionals for interviews using a combination of convenience, snowball, and theoretical sampling methods. This approach resulted in a diverse sample of 25 professionals who are actively involved in the development of AI-powered systems, including those utilizing deep learning neural networks, prediction models, large language models, and computer vision technologies for facial recognition. The participants held various roles, such as data scientists, programmers, software QA/testers, designers, and software project managers. In line with the emphasis on the importance of diversity in addressing bias in software systems and fairness debt, as highlighted by prior research [19, 37], we ensured the inclusion of individuals from equity-deserving groups, including non-male professionals, individuals with disabilities, non-white individuals, LGBTQIA+ individuals, and neurodivergent participants. This diversity was necessary for understanding different perspectives and experiences in AI development. Furthermore, 65% of the participants held advanced degrees beyond a Bachelor's, and 48% had over five years of experience in software development, indicating that the sample includes experienced professionals with substantial familiarity with the challenges of AI development.

Table 2. Demographics

| Category | Subcategory | Count |
|---|---|---|
| **Gender** | Men | 16 individuals |
| | Women | 8 individuals |
| | Non-binary | 1 individual |
| **Role** | Data Scientists | 10 individuals |
| | Designers | 5 individuals |
| | Programmers | 3 individuals |
| | Testers | 4 individuals |
| | Researchers | 2 individuals |
| | Managers | 1 individual |
| **Education** | Bachelor's Degree | 9 individuals |
| | Postbaccalaureate | 4 individuals |
| | Master's Degree | 7 individuals |
| | PhD Degree | 5 individuals |
| **Experience** | 1–3 years | 3 individuals |
| | 3–5 years | 11 individuals |
| | 5–10 years | 6 individuals |
| | More than 10 years | 5 individuals |
| **Ethnicity** | White | 20 individuals |
| | Mixed-race | 4 individuals |
| | Black | 1 individual |
| **Disability** | Without | 23 individuals |
| | With | 2 individuals |
| **LGBTQIA+** | No | 19 individuals |
| | Yes | 6 individuals |
| **Neurodivergent** | No | 21 individuals |
| | Yes | 4 individuals |

### 4.2 Indicators of Bias

The central theme that emerged from our analysis concerns recurring indicators of bias, understood as socio technical patterns that indicate underlying fairness risks in AI system development. Rather than focusing on discrete bias instances or fairness metrics, these indicators act as early signals of systemic issues that, when accumulated, can shape software outcomes and produce adverse societal effects. They point to issues across team dynamics, requirements, design processes, data handling, and implementation practices. For this study, an indicator is operationalized as a recurring socio-technical pattern observed across multiple interviews that co-occurs with fairness-relevant problems in requirements, data, modeling, evaluation, or team practices. Across our analysis, where appropriate, these patterns were corroborated by field notes and project documents.

These indicators share similarities with code smells, which signal potential problems in software code, and community smells, which expose dysfunctional patterns within development teams. Their distinct focus is on fairness risks in AI systems and the associated societal harms, from misleading outputs to reinforcement of discrimination and exclusion. The indicators identified in this study can be grouped as technical or human related. The categories reported here are preliminary rather than exhaustive and may be expanded as more projects are studied, in the same way that catalogs of code smells and community smells have evolved in prior research.

#### 4.2.1 *Technical indicators.*

Technical indicators originate from decisions made during the engineering of AI systems, including activities such as data preparation, model development, evaluation, and testing. Rather than representing fairness problems themselves, these indicators correspond to technical choices or development practices that may create opportunities for biases to remain hidden throughout the software lifecycle. Across the case, these indicators are demonstrated as decisions that are often considered routine from a software engineering perspective but, when overlooked, ignored, or deferred to later stages of development, may allow biases to become embedded in datasets, models, and other software artifacts. As these issues accumulate over successive development activities, they become progressively more difficult to identify, correct, and validate, potentially contributing to fairness related concerns after deployment. The technical indicators identified in this study are *Context Oversimplification*, *Dataset Imbalance*, *Metrics Inadequacies*, and *Ad hoc Testing*.

**Context Oversimplification** was described by participants as an indicator of bias that emerges when AI systems are developed from simplified representations of the intended operational context. Within the case under study, this occurred when relevant contextual factors were excluded or when the training data did not adequately represent the diversity of real world conditions. Participants associated this indicator with three recurring situations: limited capture conditions and environments, insufficient task and domain diversity, and limited linguistic or cultural variation. Practitioners recognized this indicator by comparing the assumptions made during development with the characteristics of the intended environment. They referred to the diversity of training data, the representation of different demographic groups, and differences in model performance across environments as common signs. Several participants also noted that these issues were often considered minor technical decisions. However, when left unaddressed, they became embedded in datasets and models, making them progressively more difficult to identify and correct. The interviews provided several examples of this indicator in practice. Participants described reducing variation in training images by using black and white images, relying on controlled lighting and backgrounds, and training models on culturally homogeneous datasets. According to their accounts, these decisions reduced the ability of the resulting models to operate consistently in more diverse real world settings. The quotations below illustrate how practitioners described this indicator in their own projects.

> **P14:** *"You will always provide your model with images of people against a white background or a specific color that contrasts with their clothing and skin tone. By doing this, you reduce complexity. For instance, your system, for now, won't be able to recognize videos of people in a crowded shopping mall or other busy environments."*
>
> **P12:** *"So I think the main problem is developing a project with a limited amount of data, doing all the training with that small dataset. Then, when you apply it in practice, it doesn't scale."*
>
> **P13:** *"So, if I make everything black and white, the system will work, but this could become a problem in the future, right?"*

**Dataset Imbalance** was identified by participants as an indicator of bias associated with unequal representation in the training data. Within the case under study, this indicator emerged when certain classes, demographic groups, or data categories were underrepresented, leading models to perform differently across groups. Participants explained that these imbalances were often difficult to recognize because models could achieve satisfactory overall performance while performing poorly for underrepresented classes. Practitioners described this indicator as becoming visible when looking beyond aggregate performance measures. They emphasized the importance of considering the distribution of training samples and comparing model performance across classes and demographic groups. According to their accounts, overlooking these imbalances during development allowed them to become embedded in the resulting models, making them progressively more difficult to identify and mitigate. The interviews provided examples of this indicator in practice. Participants described projects in which some classes were represented by considerably fewer examples than others, making it difficult for the models to learn minority cases adequately. They also reported situations in which high overall accuracy concealed poor performance for underrepresented groups, creating a misleading perception of model quality. The quotations below illustrate how practitioners described this indicator in their own projects.

**P10:** *"In our implementation, we don't always have balanced data available. This leads to errors because we are using unbalanced datasets."*

**P14:** *"But if your dataset is unbalanced, with more samples from one class than another, you might train your model and see a high accuracy of ninety percent, thinking it's performing well. However, when you look closer at the results, you may find that the model has completely misclassified an entire class."*

**P05:** *"Then, you should take a closer look at the data to assess the class balance of the problem. Depending on the dataset, it might be highly imbalanced, which could introduce bias into the model."*

**Metrics Inadequacies** was mentioned by participants as an indicator of bias associated with selecting or interpreting evaluation metrics that do not adequately reflect model performance. Within the case under study, this indicator emerged when evaluation relied primarily on aggregate measures while overlooking differences across data distributions, contexts, or demographic groups. According to participants, this often created a misleading perception of model quality and masked biases that would only become apparent in practice. Participants described metric selection as more than a technical decision, emphasizing that the choice of evaluation measures influences which problems become visible during development. They highlighted the importance of selecting metrics that reflect both the objectives of the model and the characteristics of the data, rather than relying on a single performance indicator. According to their accounts, inadequate metric selection allowed potential biases to remain unnoticed and become more difficult to identify after deployment. The interviews and observations provided examples of this indicator in practice. Participants described situations where overall accuracy was considered sufficient despite poor performance in specific scenarios. Discussions observed throughout the case also reflected concerns about the choice of evaluation criteria and the tradeoffs between different quality attributes. The quotations below illustrate how practitioners described this indicator in their own projects.

**P14:** *"But we might fall into the trap of forgetting to adjust our metrics for our specific data scenario. For example, using accuracy and claiming that our model is perfect, only to find out that it doesn't recognize anything correctly when applied, so metrics can also be a place where we unintentionally allow biases to slip through or even introduce new biases."*

**P04:** *"What I recognize is that the time spent upfront to determine the source of the data that will generate the response is extremely important for both accuracy and the quality of the information, ensuring it is reliable."*

**Field Notes [Jun 25, 2024]:** *"Participants of a machine learning group on Slack discussed whether efficiency should be valued as much as accuracy and how AI can help reduce the carbon footprint."*

**Ad-hoc Testing** emerged as an indicator of bias associated with the absence of systematic testing during AI development. Within the case under study, this indicator became apparent when teams relied on informal or opportunistic validation approaches instead of structured testing practices, particularly when fairness testing procedures were unavailable or not incorporated into the development process. Practitioners explained that these situations increased the likelihood that biases in the data or the model would remain unnoticed. Discussions throughout the projects highlighted a number of signs associated with this indicator. These included limited diversity in test data, the absence of fairness oriented evaluations, and testing procedures that did not adequately reflect the intended operational context. Interviewees also pointed out that evaluating models with data that closely resembled the training set could create a misleading perception of model quality by masking generalization problems. In their view, these practices created opportunities for biases to remain hidden and accumulate until later stages of development, where they became more difficult to identify and mitigate. Experiences within the projects also illustrated how this indicator manifested in practice. Practitioners explained that fairness related testing was frequently postponed or overlooked, particularly when organizations lacked explicit policies or guidelines for evaluating diversity and inclusion. The quotations below illustrate how this indicator was discussed throughout the case.

**P16:** *"The goal is generalization. If I'm testing with the same data I used for training, then it's not training—it's memorization."*

**P06:** *"People haven't defined a testing approach yet because it's still a new field. Development is one aspect, and testing happens later in the process. This was a common issue in earlier development practices."*

**P09:** *"If the testing phase does not include an analysis of fairness, diversity, and inclusion, and the company lacks policies to prevent software bias, evaluating the solution becomes challenging. For instance, issues such as missing fields for social identification in forms can easily be overlooked if the company does not adopt policies or strategies to emphasize software diversity."*

*4.2.2 Human related indicators.* Human related indicators originate from the people involved in AI development rather than from the technical artifacts they produce. They emerge through team interactions, decision making processes, individual assumptions, and the perspectives brought into the project. Throughout the case under study, practitioners described situations where the way teams perceived users, interpreted requirements, collaborated, and reasoned about fairness influenced development decisions. When these perspectives were narrow, homogeneous, or insufficiently questioned, potential biases could remain unnoticed and become embedded in subsequent technical artifacts, including datasets, models, requirements, and evaluation procedures. The human related indicators identified in this study are *Individual Diversity Unawareness* and *Homogeneous Team Composition*. Although closely related, they capture different aspects of AI development. The former concerns how developers understand and represent the diversity of users and stakeholders throughout the software lifecycle. The latter focuses on the diversity of perspectives, experiences, and backgrounds within the development team itself, and how these influence the team's ability to recognize assumptions, challenge decisions, and identify potential sources of bias.

**Individual Diversity Unawareness** denotes an indicator of bias that emerges when the diversity of people who use or are affected by an AI system is not adequately considered during development. Within the case under study, this indicator emerged when design and data decisions implicitly assumed a homogeneous user population rather than one shaped by differences in demographic background, culture, language, or socioeconomic circumstances. According to participants, this often resulted from datasets and requirements that did not adequately represent the intended audience. Signs of this indicator emerged in discussions about how, and to what extent, diversity was incorporated throughout development. Participants referred to the representation of different user groups in the training data, the explicit consideration of fairness during model development, and the diversity of individuals involved in evaluation as important indicators. Labeling also emerged as a recurring concern. Because labels are assigned by individuals, participants noted that they may inadvertently reflect the annotators' assumptions and perspectives, allowing biases to enter the dataset unnoticed and remain embedded throughout subsequent development activities. Practitioners' accounts illustrated this indicator in several ways. Some described understanding the characteristics of the intended users as essential for developing AI systems that perform appropriately across the target population. Others raised concerns that AI models trained on insufficiently diverse data could reproduce dominant cultural perspectives instead of adequately representing their users. The quotations below illustrate how practitioners described this indicator in their own projects.

> **P02:** *"This was my first real encounter with artificial intelligence, where I truly grasped the importance of understanding who I am working for. It's essential to understand my audience, their potential differences, and the diversity within that group to effectively train the algorithm."*
>
> **P03:** *"My current project is focused on addressing a critical issue: generative AI often perpetuates imperialistic knowledge, leading to products and services that can be racist, homophobic, and exhibit other forms of bias."*
>
> **P12:** *"So, the engineer applies this labeling, which could be based on their perception—either subjective or objective."*

*4.2.3 Homogeneous Team Composition.* represents an indicator of bias associated with limited diversity among the individuals responsible for developing AI systems. Throughout the case under study, this indicator emerged when development teams were composed of individuals with similar demographic backgrounds, professional experiences, or lived experiences, reducing the range of perspectives available during design, implementation, and evaluation. According to participants, this often limited the team's ability to recognize potential biases, particularly those affecting groups not represented within the development process. Discussions about this indicator emphasized that team diversity extends beyond demographic representation. Participants referred to differences in professional background, domain expertise, and lived experience as complementary perspectives that contribute to identifying issues that might otherwise remain unnoticed. They also noted that homogeneous teams were less likely to question assumptions embedded in datasets, requirements, and development decisions, creating opportunities for biases to remain hidden and accumulate throughout the project. Practitioners' accounts illustrated this indicator from multiple perspectives. Some emphasized that lived experience plays an important role in recognizing blind spots that may not be evident to others. Others argued that diverse teams foster broader discussions, challenge assumptions more effectively, and improve the ability to anticipate how AI systems may affect different groups of users. Participants also observed that, when relevant perspectives were absent from the development team, understanding and addressing the needs of underrepresented populations often required substantially greater effort later in the project. The quotations below illustrate how practitioners described this indicator in their own projects.

> **P4:** *"Often, people who haven't experienced a specific context may lack the empathy, sensitivity, or thoroughness needed to check, test, and analyze whether certain aspects are being overlooked or if there's a risk of bias."*
>
> **P13:** *"But I truly believe that having a diverse team—comprising people from various backgrounds, different niches, and varying perspectives—can greatly expand possibilities."*
>
> **P01:** *"So, if our team lacks LGBT individuals, Black people, Indigenous people, or people from other diverse backgrounds, important perspectives may be missed. For example, in computer vision, if we don't include these individuals in the development process, the effort and cost to understand and address their needs will be much greater."*
>
> **P18:** *"A team without diversity is more likely to produce artifacts with cognitive biases. Conversely, a team with greater diversity in gender, race, sexual orientation, and age is less likely to introduce these types of biases."*

### 4.3 What indicators do practitioners recognize as signaling potential sources of fairness debt in AI-enabled software projects?

Across the case under study, fairness was a recurring concern among AI development teams. Participants recognized that biases introduced during software development could remain unnoticed, become embedded in technical artifacts, and ultimately affect the people interacting with AI systems. These biases may contribute to unfair or discriminatory outcomes that disproportionately affect marginalized or underrepresented groups (P03, P08, P15, P17). Once embedded in AI systems, such biases often require additional auditing and accountability mechanisms to be identified and mitigated (P03, P08, P15, P17). Overlooking these issues frequently leads to increased development and maintenance costs, requiring activities such as model re-evaluation, fine-tuning, retraining, or machine unlearning (P07, P08, P12, P13, P17). Unresolved biases may also reduce opportunities for innovation and encourage the adoption of unnecessarily complex technological solutions, increasing development effort while limiting the ability of AI systems to adequately address the needs of diverse users and contexts (P02, P07, P08, P12, P14, P15, P21).

In this sense, our findings revealed six recurring indicators of bias emerging from the experiences of software professionals: *Context Oversimplification*, *Dataset Imbalance*, *Metrics Inadequacies*, *Ad hoc Testing*, *Individual Diversity Unawareness*, and *Homogeneous Team Composition*. These indicators capture how practitioners recognize potential sources of fairness debt before unfair outcomes become visible. Four indicators were associated with technical decisions involving data, models, evaluation, and testing, whereas two reflected human aspects related to understanding users, team composition, collaboration, and decision making throughout development. Our findings suggest that practitioners recognize these recurring indicators throughout the software development lifecycle. Rather than indicating that an AI system is already unfair or discriminatory, the indicators point to situations where biases may remain hidden, become embedded in software artifacts, and accumulate over time if they are overlooked or left unaddressed. Recognizing these indicators early provides development teams with opportunities to investigate, discuss, and mitigate potential sources of fairness debt before they evolve into more complex technical and societal challenges.

## 5 Discussion

In this section, we discuss our findings in light of the existing literature, and drawing on the concepts of code smells and community smells as established indicators of technical and social debt, we interpret the identified indicators of bias within the perspective of fairness debt.

Code smells and community smells have become well established concepts in software engineering for identifying potential sources of technical and social debt before they materialize into more difficult problems. Rather than representing defects themselves, they function as observable indicators that draw developers' attention to conditions requiring further investigation. Code smells point to design and implementation decisions that may compromise maintainability, whereas community smells reveal organizational and collaboration patterns that can hinder team effectiveness [9, 12, 17, 25, 46]. The analogy we draw is straightforward. If technical debt can be recognized through code smells and social debt through community smells, then fairness debt should also present observable indicators that practitioners can recognize during software development. Our findings suggest that this is indeed the case. Instead of becoming visible only after deployment through discriminatory or unfair outcomes, practitioners described recurring situations where biases could remain hidden during development and gradually accumulate into fairness debt.

A similar parallel can be observed in the origin of these indicators. Code smells typically emerge from design shortcuts, rushed implementation, or insufficient refactoring, increasing software complexity and future maintenance effort [31, 32, 36]. Community smells originate from organizational structures and collaboration practices that gradually deteriorate communication, coordination, and knowledge sharing within software teams [9, 15]. Likewise, the indicators identified in this study emerged from recurring technical and human decisions made throughout AI development. Technical indicators reflected choices concerning data, models, evaluation, and testing, whereas human-related indicators originated from how practitioners understood users, collaborated, and incorporated different perspectives into development activities. Building on these parallels, we propose referring to these observable indicators as *bias smells*:

> We define *bias smells* as observable socio-technical indicators of development decisions, practices, and team dynamics that signal situations where underlying biases may remain hidden and, if left unaddressed, contribute to the accumulation of fairness debt.

Similar to code smells and community smells, they do not imply that a system is already unfair. Instead, they signal situations where biases may remain unnoticed, become embedded in software artifacts and development practices, and progressively accumulate into fairness debt if left unaddressed. Table 3 summarizes the parallel between bias smells, code smells, and community smells.

Table 3. Summary of Characteristics, Debt, and Common Examples of Smells

| Smell | Characteristics | Common Examples |
|---|---|---|
| **Code Smells** | Technical issues such as complex code patterns and design flaws, complicating maintenance and readability. Related to technical debt. | Duplicate code, Dead code, Long parameter list, Large class, Shotgun surgery. |
| **Community Smells** | Social and organizational issues from poor socio-technical decisions, affecting team dynamics and communication. Related to social debt. | Black cloud, Cognitive distance, DevOps clash, Dispersion, Dissensus, Power distance, Organizational silo. |
| **Bias Smells** | Bias-related issues impacting technical performance and fairness, revealing deficiencies in design and team diversity. Related to fairness debt. | Context Oversimplification, Dataset Imbalance, Metrics Inadequacies, Ad hoc Testing, Individual Diversity Unawareness, and Homogeneous Team Composition. |

An important characteristic of bias smells is that they should not necessarily be interpreted as isolated indicators. Our findings suggest that technical and human-related smells may interact throughout software development. For example, limited awareness of user diversity or a homogeneous set of perspectives within a development team may influence how requirements are interpreted, which data are collected, how operational contexts are represented, and which evaluation criteria are considered appropriate. These situations may, in turn, be associated with technical smells such as Context Oversimplification, Dataset Imbalance, or Metrics Inadequacies. Bias smells should therefore be understood from a socio-technical perspective, in which human and technical conditions may interact and reinforce one another as fairness-related risks develop across software development activities.

From a software quality perspective, fairness debt extends the concerns traditionally associated with technical and social debt. Technical debt primarily affects software quality attributes such as maintainability and evolvability [25, 31, 32, 36], whereas social debt influences collaboration, coordination, communication, and overall team effectiveness [9, 12, 15, 17, 46]. Fairness debt also affects software quality, but its consequences extend beyond the development organization because AI systems directly influence decisions affecting individuals and communities [18, 38, 39]. As reported by practitioners, unresolved indicators may contribute to unfair outcomes, increased development and maintenance costs, reduced opportunities for innovation, and technology choices that inadequately address the needs of diverse populations.

By introducing *bias smells*, we extend the notion of software smells to fairness in AI-based software systems. Similar to how code smells draw attention to potential problems in software artifacts and community smells signal issues in team interactions and organizational dynamics, bias smells flag situations where biases may remain hidden and accumulate into fairness debt, eventually manifesting as issues that diminish software quality and negatively affect the people and communities interacting with these systems. From this perspective, we expect the concept of *bias smells* to contribute to the operationalization of fairness within software quality assurance by providing observable indicators that can support reviews, testing activities, and the development of engineering guidelines. However, it is important to recognize that the concepts of technical debt and code smells, as well as social debt and community smells, have evolved through decades of empirical and theoretical research that have refined their definitions and expanded the set of observable indicators. The indicators presented in this study should therefore be interpreted as preliminary. They reflect what could be identified within the scope of the investigated case and its associated data sources.

Finally, we should highlight that some of the individual indicators identified in this study have already been discussed in previous research. For example, concepts related to *Context Oversimplification* appear in requirements engineering studies [49], whereas work on software fairness has investigated challenges associated with evaluation metrics and testing [4, 16, 42, 45, 52]. Our contribution is therefore not to claim that the underlying phenomena represented by each bias smell are entirely new, but to conceptualize them collectively as observable indicators of potential sources of fairness debt. By integrating these previously disconnected observations under the notion of *bias smells* and grounding them in practitioners' experiences, we provide a common socio-technical perspective for understanding how fairness-related risks can be recognized during software development, before they manifest as discriminatory or inequitable outcomes.

## 6 Implications

This study has implications for academic research, software engineering practice, and society. By introducing the concept of *bias smells* and grounding it in practitioners' experiences, we provide a new perspective for recognizing potential sources of fairness debt throughout AI enabled software development. The following sections discuss how these findings contribute to future research, software quality assurance practices, and broader societal discussions on AI fairness.

### 6.1 Implications for Academic Research

From an academic perspective, this study contributes to the emerging body of research on software fairness by introducing the notion of *bias smells* as observable indicators that flag situations where underlying biases may remain hidden during the development of AI systems. Rather than focusing exclusively on datasets, models, or fairness metrics, our findings reinforce the view that fairness is a socio technical concern shaped by technical decisions, development practices, and human factors throughout the software development lifecycle.

This work also establishes a conceptual bridge between bias smells, code smells, and community smells. Similar to how code smells and community smells support the early identification of technical and social debt, respectively, bias smells provide a perspective for investigating potential sources of fairness debt before they become embedded in AI enabled software systems. This extends current research by complementing approaches that detect, measure, and mitigate bias in datasets, models, and system outcomes with a practitioner-oriented perspective on recognizing development conditions that may contribute to fairness debt.

Our findings also provide an empirical perspective grounded in practitioners' experiences. While previous studies have predominantly investigated fairness from algorithmic, mathematical, or ethical viewpoints, this study characterizes how software professionals recognize potential sources of fairness debt during everyday development activities. Building on these findings, our work opens several avenues for software engineering research, including the identification of additional bias smells across different AI development contexts, the investigation of how individual bias smells emerge, interact, and accumulate into fairness debt over time, and the identification of software engineering practices that are most effective for preventing or mitigating their effects. Future research can also develop and evaluate engineering guidelines and automated tool support for identifying and managing bias smells during software development. Investigating how bias smells relate to concepts such as technical debt, social debt, software quality, software process improvement, and other software engineering theories may further contribute to a broader understanding of fairness throughout the software development lifecycle.

### 6.2 Implications for Industry Practice

For practitioners, bias smells offer a practical perspective for incorporating fairness into existing software quality assurance activities. In practice, bias smells can complement established engineering processes by drawing attention to where biases may become embedded in the software development. By making these indicators more visible, we can support earlier discussions and interventions, reducing the effort required to address fairness debt after deployment. Based on the experiences shared by practitioners, we recommend the following practices to strengthen software quality assurance for AI-enabled software systems:

(1) **Integrate Bias Smells into software quality assurance.** Incorporate reviews of bias smells throughout the software lifecycle and translate them into practical quality assurance artifacts, such as review checklists, testing protocols, and engineering guidelines, to support the early identification of potential sources of fairness debt.
(2) **Systematically review data, evaluation, and testing practices.** Regularly assess the representation of operational contexts and user groups in datasets, the adequacy of evaluation metrics, and the coverage and diversity of testing strategies to identify conditions associated with context oversimplification, dataset imbalance, metrics inadequacies, and ad-hoc testing.
(3) **Incorporate diverse perspectives and fairness awareness into development.** Engage professionals with diverse backgrounds and experiences in requirements, data preparation, model design, and evaluation, while promoting ongoing fairness awareness among development teams to reduce risks associated with individual diversity unawareness and homogeneous team composition.

More broadly, our proposed bias smells provide practitioners with a practical vocabulary and a structured set of observable indicators that can support earlier identification, discussion, and mitigation of potential sources of fairness debt.

### 6.3 Implications for Society

The implications of this work also extend beyond software development organizations. As AI enabled software increasingly supports decisions in domains such as healthcare, education, finance, employment, and public services, biases introduced during development may ultimately affect individuals and communities interacting with these systems. By introducing the concept of bias smells, this study also provides a common vocabulary for discussing fairness before discriminatory outcomes become visible. Similar to how code smells facilitate discussions about software quality, bias smells offer a structured way for software engineers, managers, auditors, policymakers, and other stakeholders to discuss potential sources of unfairness and discrimination using observable situations rather than abstract ethical principles alone. However, although identifying these smells does not guarantee that unfair outcomes will be avoided, recognizing them early creates opportunities for discussion, reflection, and mitigation before biases become embedded in AI enabled software systems and potentially affect society.

## 7 Conclusion

This study investigated how software practitioners recognize potential sources of bias during the development of AI enabled software systems. Through an empirical case study involving four AI projects within a single organization, we identified six recurring indicators that practitioners associated with situations where biases may remain hidden and gradually accumulate into fairness debt during software development: *Context Oversimplification*, *Dataset Imbalance*, *Metrics Inadequacies*, *Ad hoc Testing*, *Individual Diversity Unawareness*, and *Homogeneous Team Composition.*

Based on the concepts of code smells and community smells, we introduced the notion of *bias smells*. Our findings suggest that, similar to technical debt and social debt, potential sources of fairness debt can be associated with observable socio-technical indicators arising from both technical decisions and human aspects of software development. Bias smells do not represent fairness problems themselves. Rather, they act as early warning signs that warrant further investigation and mitigation, helping practitioners identify situations where underlying biases may remain hidden and, if overlooked, accumulate into fairness debt.

Our findings contribute to research, practice, and society by introducing the notion of *bias smells* as a practical perspective for recognizing potential sources of fairness debt during AI enabled software development. By establishing a conceptual bridge between fairness debt, technical debt, and social debt, this work provides an initial foundation for incorporating fairness into software quality assurance and supporting more systematic discussions about fairness throughout the software development lifecycle. However, it is important to recognize that, unlike technical debt and social debt, whose concepts have been refined through decades of empirical research, the study of fairness debt and its observable indicators is still in its early stages. Consequently, our immediate future work will focus on validating and refining the proposed bias smells across different organizations, domains, and AI technologies using more generalizable empirical methods, including surveys, multiple case studies, and quantitative analyses.

## Data Availability

The interview guide, demographics survey, and an extended version of our chain of evidence table with quotations from participants are available at https://figshare.com/s/1adca809fe9c214090bd.

## Declaration of generative AI and AI-assisted technologies

During the preparation of this work, the authors used AI tools to assist with minor tasks such as fixing typographical issues and improving writing clarity. All content, including data collection, analysis, and interpretation, was conducted by the authors. After using this tool, the authors reviewed and edited the content as needed and take full responsibility for the content of the published article.

## Acknowledgments

This work was supported by the Natural Sciences and Engineering Research Council of Canada (NSERC), Discovery Grant RGPIN-2024-06260, and by Alberta Innovates through the Advance Program, Project Number 252608180.